\documentclass[prd,twocolumn,groupedaddress,showpacs,nofootinbib]{revtex4}
\usepackage{graphicx}
\usepackage{dcolumn}
\usepackage{amssymb}
\usepackage{mathrsfs}
\usepackage{amsmath}
\usepackage{epsfig}
\usepackage[dvips]{color}
\usepackage{hhline}

\begin{document}

\title{Leptogenesis parametrized by lepton mass matrices}

\author{Pei-Hong Gu$^1$}
\email{peihong.gu@sjtu.edu.cn}

\author{Xiao-Gang He$^{1,2,3}$}
\email{hexg@sjtu.edu.cn}

\affiliation{$^1$Department of Physics and Astronomy, Shanghai Jiao Tong
University, 800 Dongchuan Road, Shanghai 200240\\
$^2$Department of Physics, National Taiwan University, Taipei 106\\
$^3$National Center for Theoretical Sciences, Hsinchu 300}

\begin{abstract}

The conventional seesaw-leptogenesis can simultaneously explain the suppression of neutrino masses and the generation of cosmic baryon asymmetry, but usually cannot predict an unambiguous relation between these two sectors. In this work we shall demonstrate a novel left-right symmetric scenario, motivated to solve the strong CP problem by parity symmetry, where the present baryon asymmetry is determined by three charged lepton masses and a seesaw-suppressed hermitian Dirac neutrino mass matrix up to an overall scale factor. To produce the observed baryon asymmetry, this scenario requires that the neutrinos must have a normal hierarchical mass spectrum and their mixing matrix must contain a sizable Dirac CP phase. Our model can be tested in neutrino oscillation and neutrinoless double beta decay experiments.

\end{abstract}

\pacs{98.80.Cq, 14.60.Pq, 12.60.Cn, 12.60.Fr}

\maketitle

\section{Introduction}

No significant amount of primordial antimatter is found in the present universe and  any initial matter-antimatter asymmetry would have been eliminated by inflation. These facts lead to the problem of baryon asymmetry in the universe (BAU). We need a baryogenesis mechanism to dynamically generate this BAU. There have been a number of successful baryogenesis scenarios, all of which require new physics beyond the minimal standard model (SM). Furthermore, the phenomena of neutrino oscillations implies that three neutrinos should have different masses and mix with each other. This again calls for new physics since the neutrinos are massless in the SM.

It would be an interesting progress if the generation of BAU and the origin of small neutrino masses can be simultaneously understood in a unified framework. This actually has been achieved by the so-called seesaw \cite{minkowski1977} and leptogenesis \cite{fy1986} schemes where the lepton number is violated by two units and the neutrinos are Majorana particles. One should keep in mind, however,  that the Majorana nature of neutrinos has not been confirmed by any experiments. It has been shown that even if the neutrinos turn out to be Dirac particles, one can realize the seesaw and leptogensis mechanisms by introducing additional symmetries besides some heavy degrees of freedom to the theory \cite{rw1983,dlrw1999,gh2006}. Compared with Majorana neutrino models, Dirac neutrino models usually do not violate the lepton number, but need more unknown parameters. In general, for both of Majorana and Dirac seesaw-leptogenesis scenarios, one does not know much about the texture of the masses and couplings involving the new heavy fields, and hence cannot give an unambiguous relation between the BAU and the information reside in the neutrino mass matrix, such as  mass hierarchy and CP phase(s). The predictive power is quite limited \cite{di2001}.

In this paper we shall develop a novel scenario where the Dirac neutrinos have a seesaw-suppressed hermitian mass matrix and up to an overall scale factor, the baryon asymmetry is determined by the neutrino and charged lepton mass matrices. To construct a realistic model we also aim at the strong CP problem which is another big challenge to the SM. Specifically we shall consider an $SU(3)^{}_{c}\times SU(2)^{}_{L}\times SU(2)^{}_{R}\times U(1)^{}_{L}\times U(1)^{}_{R}\rightarrow SU(3)^{}_{c}\times SU(2)^{}_{L}\times SU(2)^{}_{R}\times U(1)^{}_{B-L}$ left-right symmetric \cite{ps1974} model with a softly broken parity symmetry to solve the strong CP problem without an unobserved axion \cite{ms1978,bm1989}. Our leptogenesis also requires the existence of a lepton number violation which, however, is not allowed to contribute any Majorana neutrino masses. For this purpose we shall impose an unbroken $Z_4$ discrete symmetry. To produce the BAU, the neutrino mass matrix is required to satisfy a normal hierarchy pattern and contain a sizable Dirac CP phase. Our model can be tested in the running and forthcoming neutrino oscillation experiments such as NO$\nu$A and JUNO, and can be excluded by any observations of neutrinoless double beta decay.

\section{The model}

In this model three generations of fermions and their corresponding $SU(3)^{}_{c}\times SU(2)^{}_{L}\times SU(2)^{}_{R}\times U(1)^{}_{L}\times U(1)^{}_{R}\times Z_4$ quantum numbers are as the following,
\begin{eqnarray}
Q_L^{}(3, 2, 1, \frac{1}{3}, 0, i)&\stackrel{P}{\longleftrightarrow} &Q_R^{}(3, 1, 2, 0, \frac{1}{3}, i)\,,\nonumber\\
U_R^{}(3, 1,1,\frac{4}{3}, 0, i)& \stackrel{P}{\longleftrightarrow} & U_L^{}(3, 1, 1, 0, \frac{4}{3}, i)\,,\nonumber\\
D_R^{}(3, 1, 1, -\frac{2}{3}, 0, i) &\stackrel{P}{\longleftrightarrow}& D_L^{}(3, 1, 1, 0, -\frac{2}{3}, i)\,,\nonumber\\
L_L^{}(1,2,1,-1,0, i)&\stackrel{P}{\longleftrightarrow} &L_R^{}(1,1, 2, 0, -1, i)\,,\nonumber\\
e_R^{}(1,1,1,-2,0,i)&\stackrel{P}{\longleftrightarrow}&E_L^{}(1,1, 1, 0, -2, i)\,,
\end{eqnarray}
where the components of the $SU(2)$ doublets are
\begin{eqnarray}
&&Q_{L,R}^{}=\left[\begin{array}{c}u_{L,R}^{}\\
d_{L,R}^{}\end{array}\right],~
L_{L}^{}=\left[\begin{array}{c}\nu_{L}^{}\\
e_{L}^{}\end{array}\right],~L_{R}^{}=\left[\begin{array}{c}\nu_{R}^{}\\
E_{R}^{}\end{array}\right].
\end{eqnarray}
As we will show later that the new particles $U_{L,R}$, $D_{L,R}$ and $E_{L,R}$ are heavy.

The scalars in this model are
\begin{eqnarray}
\omega_U^{}(1, 1, 1, \frac{4}{3}, -\frac{4}{3}, 1)&\stackrel{P}{\longleftrightarrow}&\omega_U^\dagger\,,\nonumber\\
\omega_D^{}(1, 1, 1, -\frac{2}{3}, \frac{2}{3}, 1)&\stackrel{P}{\longleftrightarrow}&\omega_D^\dagger\,,\nonumber\\
\Sigma(1, 2, 2, -1, 1, 1)&\stackrel{P}{\longleftrightarrow}&\Sigma^\dagger_{}\,,\nonumber\\
\phi_L^{}(1, 2, 1, -1,0, 1)&\stackrel{P}{\longleftrightarrow}&\phi_R^{}(1, 1, 2, 0, -1, 1)\,,\nonumber\\
\Delta_L^{}(1, 3, 1, 2, 0, -1)&\stackrel{P}{\longleftrightarrow}&\Delta_R^{}(1, 1, 3, 0, 2, -1)\,, \nonumber\\
\chi(1, 1, 1, 0, 0, -1)&\stackrel{P}{\longleftrightarrow}&\chi\,.
\end{eqnarray}
Symmetry breaking is driven by nonzero vacuum expectation values (VEV) of scalars. One can write down the full renormalizable scalar potential to analyze the breaking pattern in details. Here we will, instead, write down a few of the terms which play some special roles,
\begin{eqnarray}
\label{potential}
V&\supset&\mu_{\phi_L^{}}^{2}\phi^\dagger_L\phi_L^{}+\mu_{\phi_R^{}}^{2}\phi^\dagger_R\phi_R^{}
+\zeta_1^{}\phi_R^\dagger\Sigma^\dagger_{}\Sigma\phi_R^{}\nonumber\\
&&
+\zeta_2^{}\tilde{\phi}_R^\dagger\Sigma^\dagger_{}\Sigma\tilde{\phi}_R^{}+[\rho\omega_U^{}\omega_D^2+\mu\phi_L^\dagger\Sigma\phi_R^{}\nonumber\\
&&
+\lambda\omega_U^{\dagger}\omega_D^{}\textrm{Tr}(\Sigma^\dagger_{}\tilde{\Sigma})
+\kappa \chi (\phi_L^T i\tau_2^{}\Delta_L^{}\phi_L^{}\nonumber\\
&&+\phi_R^T i\tau_2^{}\Delta_R^{}\phi_R^{})+\textrm{H.c.}]\,.
\end{eqnarray}

When the $\omega_D^{}$ scalar develops its VEV, the $U(1)_L^{}\times U(1)_R^{}$ symmetry will be broken down to a $U(1)_{B-L}^{}$, i.e. $U(1)^{}_{L}\times U(1)^{}_{R}\stackrel{\langle\omega_D^{}\rangle}{\longrightarrow} U(1)^{}_{B-L}$. Naively, if the $\omega_U^{}$ scalar takes a positive mass $M_{\omega_U}^2$, its VEV should be zero. However,
due to the $\rho$-term in the above potential, one can get an induced VEV, $\langle\omega_U^{}\rangle\simeq - \rho\langle\omega_D^{}\rangle^2_{} /M_{\omega_U}^2\leq\langle\omega_D^{}\rangle$. Subsequently,
the $\phi_R^{}$ and $\phi_L^{}$ scalars will respectively drive the left-right symmetry breaking $SU(2)_R^{}\times U(1)_{B-L}^{}\stackrel{\langle\phi_R^{}\rangle}{\longrightarrow} U(1)_{Y}^{}$ and the electroweak symmetry breaking $SU(2)_L^{}\times U(1)_{Y}^{}\stackrel{\langle\phi_L^{}\rangle}{\longrightarrow} U(1)_{em}^{}$.
In order to separate the left-right and electroweak symmetry breaking scales, we softly break the parity symmetry by taking $\mu_{\phi_L^{}}^{2}\neq \mu_{\phi_R^{}}^{2}$.
As a result of the softly broken parity, the left-right symmetry breaking scale can be much higher than the electroweak one, i.e. $\langle\phi_R^{}\rangle\gg\langle\phi_L^{}\rangle$.

After the left-right symmetry breaking, the Higgs bidoublet $\Sigma$ can be conveniently treated as two $SU(2)_L^{}$ doublets,
$\Sigma\equiv [\sigma_1^{}~\sigma_2^{}]$. The $\sigma_1^{}$ and $\sigma_2^{}$ scalars can have a large mass split because of the $\zeta_{1,2}^{}$-terms in the potential (\ref{potential}). By choosing the masses $M_{\sigma_{1,2}^{}}^2$ of $\sigma_{1,2}^{}$ positive and very large, the $\mu$-term and the $\lambda$-term in the potential will lead to the induced small VEVs for $\sigma_{1,2}$ as below,
\begin{eqnarray}
\langle\sigma_{1}^{}\rangle\simeq -\frac{\mu\langle\phi^{}_L\rangle\langle\phi^{}_R\rangle}{M_{\sigma_1^{}}^2}\geq\langle\sigma_{2}^{}\rangle\simeq -\frac{\lambda\langle\omega_D^{}\rangle\langle\omega_U^{}\rangle\langle\sigma^{}_1\rangle}{M_{\sigma_2^{}}^2}\,.~~~~
\end{eqnarray}
As we will show later a successful leptogenesis requires the VEV $\langle\sigma_{1}^{}\rangle$ below the eV scale.

We emphasize that the $Z_4^{}$ discrete symmetry will not be broken at any scales. Accordingly, the real singlet scalar $\chi$ and the $[SU(2)]$-triplet scalars $\Delta_{L,R}^{}$ will not have any nonzero VEVs and the neutrinos will not develop nonzero Majorana masses.

\section{Fermion masses}

Fermion masses can be generated at tree level through the Yukawa couplings as follows,
\begin{eqnarray}
\label{yukawa}
\mathcal{L}_Y^{}&=& -y_U^{}(\bar{Q}_L^{}\phi_L^{}U_R^{}+\bar{Q}_R^{}\phi_R^{}U_L^{})-f_U^{}\omega_U^{}\bar{U}_L^{}U_R^{}\nonumber\\
&&-y_D^{}(\bar{Q}_L^{}\tilde{\phi}_L^{}D_R^{}+\bar{Q}_R^{}\tilde{\phi}_R^{}D_L^{})-f_D^{}\omega_D^{}\bar{D}_L^{}D_R^{}\nonumber\\
&&-y_e^{}(\bar{L}_L^{}\tilde{\phi}_L^{}e_R^{}+\bar{L}_R^{}\tilde{\phi}_R^{}E_L^{})-f_\Sigma\bar{L}_L^{}\Sigma L_R^{}\nonumber\\
&&-f_\Delta(\bar{L}_L^c i\tau_2^{}\Delta_L^{} L_L^{}+\bar{L}_R^c i\tau_2^{}\Delta_R^{} L_R^{})+\textrm{H.c.}\,.
\end{eqnarray}
The parity symmetry, although softly broken, still enforce $f_{U(D,\Sigma)}^{}=f_{U(D,\Sigma)}^{\dagger}$.

The mass matrices of the ordinary $[SU(2)]$-doublet quarks and the new $[SU(2)]$-singlet quarks are given by
\begin{eqnarray}
\mathcal{L}&\supset&-\left[\begin{array}{ll}\bar{u}_L^{}&\bar{U}_L^{}\end{array}\right]
M_u^{}
\left[\begin{array}{c}u_R^{}\\
U_R^{}\end{array}\right]-\left[\begin{array}{ll}\bar{d}_L^{}&\bar{D}_L^{}\end{array}\right]M_d^{}
\left[\begin{array}{c}d_R^{}\\D_R^{}\end{array}\right]\nonumber\\
&&+\textrm{H.c.}~~\textrm{with}\nonumber\\
&&M_{u(d)}^{}=\left[\begin{array}{cc}0&y_{U(D)}^{}\langle\phi_L^0\rangle\\
y_{U(D)}^{\dagger}\langle\phi_R^0\rangle&f_{U(D)}^{}\langle\omega_{U(D)}^{}\rangle\end{array}\right]\,.
\end{eqnarray}
The quark mass matrices $M_{u(d)}^{}$ have real determinants and hence do not contribute to the strong CP phase $\bar{\Theta}=\Theta - \textrm{Arg}\textrm{Det}(M_u^{}M_d^{})$ at tree level. Here $\Theta$ is from the QCD $\Theta$-vacuum and should have been removed by the parity symmetry. The loop-level contribution from the quark mass matrices to the strong CP phase can be below the experimental limit when the new quarks are heavy enough \cite{bm1989}. Therefore, the strong CP problem can be solved without introducing an unobserved axion.

The lepton mass matrices are very different from those for quarks. The mass matrices for the ordinary charged leptons $e_{L,R}$ and the mirror charged leptons $E_{L,R}^{}$ are
\begin{eqnarray}
\mathcal{L}\supset-\left[\begin{array}{ll}\bar{e}_L^{}&\bar{E}_L^{}\end{array}\right]
\left[\begin{array}{cc}y_e^{}\langle\phi_L^0\rangle&f_\Sigma^{}\langle\sigma_2^0\rangle\\
0&y_e^{\dagger}\langle\phi_R^0\rangle\end{array}\right]
\left[\begin{array}{c}e_R^{}\\
E_R^{}\end{array}\right]+\textrm{H.c.}\,.~~
\end{eqnarray}
For $\langle\sigma_{2}^0\rangle\leq \langle\sigma_{1}^0\rangle\ll\langle\phi_L^0\rangle\ll\langle\phi_R^0\rangle$,
one obtains the light and heavy charged lepton mass matrices,
\begin{eqnarray}
y_e^{}\langle\phi_L^0\rangle
\propto y_e^{}\langle\phi_R^0\rangle \Rightarrow\frac{M_{E_e}^{}}{m_e^{}}=\frac{M_{E_\mu}^{}}{m_\mu^{}}
=\frac{M_{E_\tau}^{}}{m_\tau^{}}=\frac{\langle\phi_R^0\rangle}{\langle\phi_L^0\rangle}\,.
\end{eqnarray}
Without loss of generality and for convenience, we will work in the base where the above charged lepton mass matrices are real and diagonal.

As for the left-handed neutrinos $\nu_L^{}$ and the right-handed neutrinos $\nu_R^{}$, they can form Dirac pairs and acquire a hermitian mass matrix,
\begin{eqnarray}
\label{numassmatrix}
\!\!\!\!\!\!\!\!\!\!\!\!&&\mathcal{L} \supset -m_\nu^{}\bar{\nu}_L^{}\nu_R^{}+\textrm{H.c.}\,,\nonumber\\
\!\!\!\!\!\!\!\!\!\!\!\!&&m_\nu^{}=f_\Sigma^{}\langle\sigma_1^0\rangle=U\hat{m}_\nu^{}U^\dagger_{}=U\textrm{diag}\{m_1^{},m_2^{},m_3^{}\}
U^\dagger_{}\,,
\end{eqnarray}
with $U$ being the PMNS \cite{pmns1957} matrix which now does not contain any Majorana CP phases. Taking into account the perturbation requirement $|(f_\Sigma^{})_{\alpha\beta}|<\sqrt{4\pi}$, the VEV $\langle\sigma_1^0\rangle$ should have a low limit. For the neutrino mass spectrum being normal hierarchy (NH) $m_3^{}>m_2^{}>m_1$, or inverted hierarchy (IH) $m_2^{}>m_1> m_3$, we can constrain
\begin{eqnarray}
\label{pert}
\!\!\!\!\langle\sigma_1^0\rangle^2_{}&>&\frac{m_3^2}{4\pi} \geq1.96\times 10^{-4}_{}\,\textrm{eV}^2_{}~ (\textrm{NH})\,,\nonumber\\
\!\!\!\!\langle\sigma_1^0\rangle^2_{}&>&\frac{m_2^2}{4\pi}\geq 2.01\times 10^{-4}_{}\,\textrm{eV}^2_{}~(\textrm{IH})\,,
\end{eqnarray}
by inputting the best fitting values \cite{gms2014}.

\begin{figure}
\vspace{4.5cm} \epsfig{file=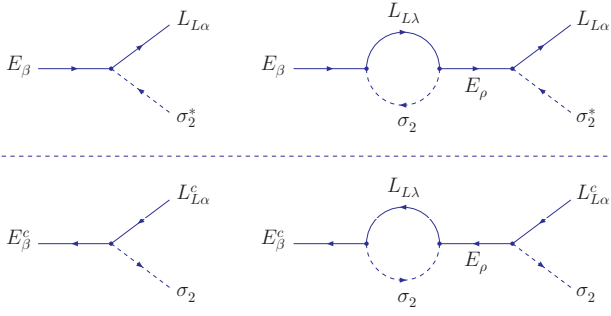, bbllx=4.8cm, bblly=6.0cm,
bburx=14.8cm, bbury=16cm, width=5.5cm, height=5.5cm, angle=0,
clip=0} \vspace{-5.9cm} \caption{\label{decay} The lepton number conserving decays of the mirror charged lepton-antilepton pairs.}
\end{figure}

\section{Lepton and baryon asymmetries}

After the left-right symmetry is broken down to the electroweak symmetry, the Yukawa interactions in Eq. (\ref{yukawa}) will result in
\begin{eqnarray}
\label{yukawa2}
\mathcal{L}\supset- f_\Sigma^{}\bar{L}_L^{}\sigma_1^{}\nu_R^{}- f_\Sigma^{}\bar{L}_L^{}\sigma_2^{}E_R^{}-\hat{M}_E^{}\bar{E}_R^{}E_L^{}+\textrm{H.c.}.
\end{eqnarray}
Therefore, the mirror charged leptons $E_{e,\mu,\tau}^{}$ can decay into the ordinary leptons $L_{L\alpha}^{}~(\alpha=e,\mu,\tau)$ and the heavy Higgs scalar $\sigma_2^{}$ if kinematically allowed. Evaluating the diagrams in Fig. \ref{decay}, we obtain the flavor-dependent CP asymmetries in the decays of the mirror lepton-antilepton pairs as
\begin{eqnarray}
\label{cpa}
\!\!\!\!\varepsilon_{\beta\alpha}^{}\!\!&=&\!\!\frac{\Gamma(E_\beta^{}\rightarrow L_{L\alpha}^{}\sigma_2^\ast)-\Gamma(E_\beta^{c}\rightarrow L_{L\alpha}^{c}\sigma_2^{})}{\Gamma_{E_\beta^{}}^{}}\nonumber\\
\!\!\!\!\!\!&=&\!\!\frac{1}{4\pi}\sum_\rho^{}
\frac{\textrm{Im}[(f^\dagger_{\Sigma}f^{}_\Sigma)_{\rho \beta}^{}(f^\dagger_{\Sigma})_{\beta\alpha}^{}(f^{}_\Sigma)_{\alpha\rho}^{}]}
{(f^\dagger_{\Sigma}f^{}_\Sigma)_{\beta\beta}^{}}\frac{M_{E_\beta}^{2}}{M_{E_\rho}^{2}-M_{E_\beta}^{2}}\nonumber\\
\!\!\!\!\!\!&=&\!\!\frac{1}{4\pi}\sum_\rho^{}
\frac{\textrm{Im}[(m^\dagger_{\nu}m_\nu^{})_{\rho\beta}^{}(m^\dagger_{\nu})_{\beta\alpha}^{}(m_\nu^{})_{\alpha\rho}^{}]}
{(m^\dagger_{\nu}m_\nu^{})_{\beta\beta}^{}\langle\sigma_1^0\rangle^2_{}}\frac{m_{\beta}^{2}}{m_{\rho}^{2}-m_{\beta}^{2}}\,.\nonumber\\
&&
\end{eqnarray}
It is easy to see that $\sum_{\alpha=e,\mu,\tau}^{}\varepsilon_{\beta\alpha}^{}=0$ and $\varepsilon_{\beta\alpha}^{}$ can be fully determined by the Dirac neutrino mass matrix $m_\nu^{}$ and the charged lepton mass matrix $\hat{m}_e^{}$ for a given VEV $\langle\sigma_1^0\rangle$. We find that the above expression is proportional to the Jarlskog parameter $J_{CP}=\textrm{Im}(U_{e1}^{}U_{e2}^{*}U_{\mu 1}^{*}U_{\mu 2}^{})$ \cite{jarlskog1985}. For example, $\varepsilon_{e\alpha}^{}~(\alpha=e,\mu,\tau)$ can be written as
\begin{eqnarray}
\label{cpae}
\varepsilon_{ee}^{}&=&-\frac{1}{4\pi}\frac{J_{CP}^{}F(\hat{m}_\nu^{})
(m_\nu^{})_{ee}^{}}
{(m^\dagger_{\nu}m_\nu^{})_{ee}^{}\langle\sigma_1^0\rangle^2_{}}\frac{m_e^2}{m_\mu^2}\,,\nonumber\\
\varepsilon_{e\mu}^{}&=&-\frac{1}{4\pi}\frac{J_{CP}^{}F(\hat{m}_\nu^{})(m_\nu^{})_{\mu\mu}^{}}
{(m^\dagger_{\nu}m_\nu^{})_{ee}^{}\langle\sigma_1^0\rangle^2_{}}\frac{m_e^2}{m_\mu^2}\,,\nonumber\\
\varepsilon_{e\tau}^{}&=&-(\varepsilon_{ee}^{}+\varepsilon_{e\mu}^{})\,,
\end{eqnarray}
with
\begin{eqnarray}
F(\hat{m}_\nu^{})&=&m_1^{}m_2^{}(m_1^{}-m_2^{})+m_2^{}m_3^{}(m_2^{}-m_3^{})\nonumber\\
&&+m_3^{}m_1^{}(m_3^{}-m_1^{})\,.
\end{eqnarray}
Note $\varepsilon_{ee}^{}$ and $\varepsilon_{e\mu}^{}$ have the same sign and are sensitive to whether the neutrino masses satisfy the NH or IH pattern.

\begin{figure}
\vspace{4.6cm} \epsfig{file=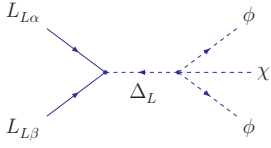, bbllx=3.8cm, bblly=6.0cm,
bburx=13.8cm, bbury=16cm, width=5.5cm, height=5.5cm, angle=0,
clip=0} \vspace{-8cm} \caption{\label{scattering} The lepton number violating annihilations of the ordinary leptons.}
\end{figure}

Although these lepton number conserving decays do not generate a net lepton number, they can produce three individual lepton asymmetries $L_{e,\mu,\tau}^{}$ stored in the ordinary electron, muon and tau lepton flavors. Since the mirror electron is much lighter than the mirror muon and tau, the decays of the mirror electron-positron pairs can be expected to determine the individual lepton asymmetries $L_{e,\mu,\tau}^{}$. If the mirror electron and positron are heavy enough, their decays can produce the lepton asymmetries $L_{e,\mu,\tau}^{}$ before the sphaleron \cite{krs1985} processes stop working. Furthermore, the $[SU(2)_L^{}]$-triplet scalar $\Delta_L^{}$ can mediate some lepton number violating processes, such as the lepton number violating annihilation shown in Fig. \ref{scattering}. For a proper choice of the couplings and mass of the $\Delta_L^{}$ scalar, only the lepton number violating interactions of certain lepton flavor(s) can keep in equilibrium when the individual lepton asymmetries $L_{e,\mu,\tau}^{}$ are produced. For this reason a nonzero lepton asymmetry, i.e. $L_e^{}$, $L_\mu^{}$, $L_\tau^{}$, $L_e^{}+L_\mu^{}$, $L_e^{}+L_\tau^{}$ or $L_\mu^{}+L_\tau^{}$, can survive from the lepton number violating processes. If this remaining lepton asymmetry is induced before the sphaleron epoch, it can be partially converted to a baryon asymmetry.

The interaction rate of the flavor-dependent lepton number violating processes, from Fig. \ref{scattering}, is given by,
\begin{eqnarray}
\Gamma^{}_{\alpha\beta}&=&\Gamma(L_{L\alpha}^{}L_{L\beta}^{}\rightarrow \phi\phi\chi)\nonumber\\
&=&\left\{\begin{array}{lcl}\frac{3}{2^5_{}\pi^5_{}}\kappa^2_{}|(f_\Delta^{})_{\alpha\beta}^{}|^2_{}
\frac{T^5_{}}{M_\Delta^4}&\textrm{for}&T<M_\Delta^{}\,,\\
\frac{1}{2^{11}_{}\pi^5_{}}\kappa^2_{}|(f_\Delta^{})_{\alpha\beta}^{}|^2_{}
T&\textrm{for}&T>M_\Delta^{}\,.\end{array}\right.
\end{eqnarray}
As an example, we consider the case where only the $L_\tau^{}$ asymmetry can be quickly washed out once they are produced, if the following conditions are satisfied \cite{kt1990}
\begin{eqnarray}
&&(\Gamma^{}_{\tau\tau}>H)\left|_{T=M_{E_e}^{}}^{}\right.,\nonumber\\
&&(\Gamma^{}_{\alpha\beta}<H)\left|_{T=M_{E_e}^{}}^{}\right. ~~((\alpha,\beta)\neq(\tau,\tau))\,,
\end{eqnarray}
where $H=\left(8\pi^{3}_{}g_{\ast}^{}/90\right)^{\frac{1}{2}}_{}
T^{2}_{}/M_{\textrm{Pl}}^{}$ is the Hubble constant
with $M_{\textrm{Pl}}^{}\simeq 1.22\times 10^{19}_{}\,\textrm{GeV}$ being the Planck mass and $g_{\ast}^{}=117$ being the relativistic degrees of freedom (the SM fields, the real singlet scalar, the heavy Higgs doublet $\sigma_2^{}$, the right-handed neutrinos). This means that for $M_\Delta^{}>M_{E_e}^{}$, the Yukawa couplings $f_\Delta^{}$ should take the structure,
\begin{eqnarray}
|(f_\Delta^{})_{\tau\tau}^{}|^2_{}\!&>&\! 0.0048 \left(\frac{1}{\kappa}\right)^2_{} \left(\frac{10^{12}_{}\,\textrm{GeV}}{M_{E_e}^{}}\right)^{3}_{}
\left(\frac{M_\Delta^{}}{M_{E_e}^{}}\right)^{4}_{}\,, \nonumber\\
|(f_\Delta^{})_{\alpha\beta}^{}|^2_{}\!&<&\!0.0048 \left(\frac{1}{\kappa}\right)^2_{} \left(\frac{10^{12}_{}\,\textrm{GeV}}{M_{E_e}^{}}\right)^{3}_{}
\left(\frac{M_\Delta^{}}{M_{E_e}^{}}\right)^{4}_{} \nonumber\\
&&((\alpha,\beta)\neq(\tau\tau))\,,
\end{eqnarray}
which may be induced by some symmetry arrangements like the Froggatt-Nielsen mechanism \cite{fg1979}.

The mirror electron mass is above the sphaleron temperature but much below the high left-right symmetry breaking scale. Therefore, the gauge interactions of the mirror electron and positron can go out of equilibrium before the efficient decays of the mirror electron-positron pairs. In the weak washout region, the out-of-equilibrium condition of the mirror electron and positron can be simply described by $\Gamma_{E_e}^{}/(2H)\left|_{T=M_{E_e}^{}}^{}\right.<1$.
The individual lepton asymmetries induced by the decays of the mirror electron-positron pairs then can approximate to \cite{kt1990}
\begin{eqnarray}
\!\!L_\alpha^{}\!=\!\frac{n_L^{}}{s}\simeq\varepsilon_{e\alpha}^{}\frac{n_{E_e}^{\textrm{eq}}}{s}\left|_{T=M_{E_e}^{}}^{}\right.\!=\! \frac{45}{2^{\frac{3}{2}}_{}\pi^{\frac{7}{2}}_{}e} \frac{\varepsilon_{e\alpha}^{}}{g_\ast^{}}\!=\!\frac{0.11\,\varepsilon_{e\alpha}^{}}{g_\ast^{}}\,.
\end{eqnarray}
From the out-of-equilibrium condition and the neutrino oscillation data \cite{gms2014}, we can constrain
the VEV $\langle\sigma_1^0\rangle$ by
\begin{eqnarray}
&&\langle\sigma_1^0\rangle^2_{}\!\!>\! \!\frac{1}{32\pi}{T^2\over H}
\frac{(m_\nu^\dagger m_\nu^{})_{ee}^{}}{M_{E_e}^{}}\nonumber\\
\!\!&&=\!\!\left\{\begin{array}{lcl}\!\!1.96\times 10^{-4}_{}\,\textrm{eV}^2_{}\left(\frac{5.24\times  10^{15}_{} \textrm{GeV}}{M_{E_e}^{}}\right)\,,\textrm{NH with} \, m_1^{}=0\,,\\
[5mm]
\!\!2.01\times 10^{-4}_{}\,\textrm{eV}^2_{}\left(\frac{1.62\times 10^{17}_{}\textrm{GeV}}{M_{E_e}^{}}\right)\,, \textrm{IH with }\, m_3^{}=0\,,\end{array}\right.
\end{eqnarray}
which is allowed to match the perturbation condition for a large mirror electron mass $M_{E_e}^{}$.
In the above, for definitiveness we have taken $m_1 = 0$ and $m_3 =0$ for NH and IH, respectively. These choices give the lower bounds for $\langle\sigma_1^0\rangle^2_{}$, and also the upper bounds for $\varepsilon_{ee}^{}+\varepsilon_{e\mu}^{}$ which will be shown later in Eq. (\ref{eee}).

The sphaleron processes can partially convert a remaining lepton asymmetry $L_e^{}$, $L_\mu^{}$, $L_\tau^{}$, $L_e^{}+L_\mu^{}$, $L_e^{}+L_\tau^{}$ or $L_\mu^{}+L_\tau^{}$ to a baryon asymmetry. For example, we consider an $L_e^{}+L_\mu^{}$ asymmetry from the decays of the mirror electron-positron pairs to be the remaining lepton asymmetry. The final baryon asymmetry then should be \cite{kt1990}
\begin{eqnarray}
\label{bas}
\eta_B^{}&=&7.04\times \frac{n_B^{}}{s}=5.91\times 10^{-10}_{}\left(\frac{\varepsilon_{ee}^{}+\varepsilon_{e\mu}^{}}{-2.52\times 10^{-7}_{}}\right)\nonumber\\
&&\textrm{with}~~\frac{n_B^{}}{s}=-\frac{28}{79}\frac{n_L^{}}{s}\simeq -\frac{28}{79}\frac{0.11(\varepsilon_{ee}^{}+\varepsilon_{e\mu}^{})}{g_\ast^{}}\,.~~~~
\end{eqnarray}
Here the factor $-28/79$ is the lepton-to-baryon conversion coefficient \cite{ht1990}.
By inputting the known neutrino oscillation data \cite{gms2014}
into the CP asymmetries (\ref{cpae}), which do not have any cancellations between $\varepsilon_{ee}^{}$ and $\varepsilon_{e\mu}^{}$, we obtain
\begin{eqnarray}
\label{cpamm}
\varepsilon_{ee}^{}+\varepsilon_{e\mu}^{}&=&2.52\times 10^{-7}_{} \left(\frac{\sin\delta}{0.125}\right)\left(\frac{1.96\times 10^{-4}_{}\,\textrm{eV}^2_{}}
{\langle\sigma_1^0\rangle^2_{}}\right) \nonumber\\
&&\textrm{for NH with}~~m_1^{}=0\,,\nonumber\\
\varepsilon_{ee}^{}+\varepsilon_{e\mu}^{}&=&1.61\times 10^{-8}_{} \left(\frac{\sin\delta}{1}\right)\left(\frac{2.01\times 10^{-4}_{}\,\textrm{eV}^2_{}}
{\langle\sigma_1^0\rangle^2_{}}\right) \nonumber\\
&&\textrm{for IH with}~~m_3^{}=0\,. \label{eee}
\end{eqnarray}
To obtain the observed value,
$\eta_B^{}=5.91\times 10^{-10}_{}$ \cite{ade2013}, the CP asymmetry $\varepsilon_{ee}^{}+\varepsilon_{e\mu}^{}$ needs to satisfy
$(\varepsilon_{ee}^{}+\varepsilon_{e\mu}^{})\sim -2.52\times 10^{-7}_{}$.

Comparing the above with Eq. (\ref{pert}), we conclude that our model with the IH neutrinos cannot solve the BAU problem, but with the NH neutrinos can if the Dirac CP phase satisfies,
\begin{eqnarray}
\label{prediction}
-\sin\delta &\simeq& 0.125 \left(\frac{\langle\sigma_1^0\rangle^2_{}}{1.96\times 10^{-4}\,\textrm{eV}^2_{}}\right)\geq 0.125 \nonumber\\
&&\textrm{for}~~\langle\sigma_1^0\rangle^2_{}\geq 1.96\times 10^{-4}\,\textrm{eV}^2_{}\,.
\end{eqnarray}
Note that recent NO$\nu$A result \cite{bian2015} also hints the NH neutrino mass spectrum. For the currently favored value $\sin\delta \sim -1$ \cite{palazzo2015} , we get $\langle\sigma_1^0\rangle^2_{}\sim 1.6\times 10^{-3}_{}\,\textrm{eV}^2_{}$.

\section{Summary}

In this paper we have described a novel left-right symmetric scenario where the strong CP problem can be solved by parity, the neutrinos can acquire a seesaw-suppressed Dirac mass matrix, while the BAU can be determined by the Dirac neutrino and charged lepton mass matrices up to an overall scale. To solve the BAU problem our model predicts a NH pattern for the neutrino masses and a sizable Dirac CP phase in the mixing matrix. We look forward to testing our model by new data from the NO$\nu$A and JUNO, as well as neutrinoless double beta decay experiments.

\textbf{Acknowledgement}: PHG was supported by the Shanghai Jiao Tong University (Grant No. WF220407201) and the Recruitment Program for Young Professionals (Grant No. 15Z127060004). HXG was supported in part by MOE Academic Excellent Program (Grant No. 102R891505) and MOST of ROC, and in part by NSFC (Grant No. 11175115). The authors were also supported by the Shanghai Laboratory for Particle Physics and Cosmology under Grant No. 11DZ2260700.

\end{document}